\documentclass[12pt,reqno,a4paper]{amsart}
\usepackage{amsmath,amssymb,amsfonts,amscd}
\usepackage[mathscr]{eucal}
\usepackage{comment}
\usepackage{hyperref}

\date{28 September (11 October)   2013}

\author{Theodore Th. Voronov}
\address{School of Mathematics, University of Manchester,
Oxford Road, Manchester, M13 9PL, United Kingdom}
\email{theodore.voronov@manchester.ac.uk}
\title[A converse to the AKSZ construction]{Vector fields on mapping spaces and a converse to the AKSZ construction}

\newtheorem{thm}{Theorem}
\newtheorem{prop}{Proposition}
\newtheorem{cor}{Corollary}
\newtheorem*{coro}{Corollary}
\theoremstyle{definition}
\newtheorem{ex}{Example}
\newtheorem{rem}{Remark}

\def\co{\colon\thinspace}

\DeclareMathOperator{\Hom}{Hom}
\DeclareMathOperator{\Homm}{\mathbf{Hom}}

\DeclareMathOperator{\Mapp}{\mathbf{Map}}
\DeclareMathOperator{\Map}{{Map}}

\DeclareMathOperator{\Diff}{Diff}

\newcommand{\der}[2]{{\frac{\partial {#1}}{\partial {#2}}}}
\newcommand{\oder}[2]{{\frac{d {#1}}{d {#2}}}}

\newcommand{\var}[2]{{\frac{\delta {#1}}{\delta {#2}}}}
\newcommand{\lvar}[2]{{{\delta {#1}}/{\delta {#2}}}}

\newcommand{\R}[1]{{\mathbb R}^{#1}}

\newcommand{\p}{\partial}

\renewcommand{\a}{\alpha}
\renewcommand{\b}{\beta}
\newcommand{\e}{\varepsilon}

\newcommand{\f}{{\varphi}}
\renewcommand{\O}{\Omega}

\renewcommand{\o}{\omega}

\newcommand{\h}{\eta}

\newcommand{\la}{{\lambda}}

\renewcommand{\d}{\delta}

\newcommand{\itt}{{\tilde \imath}}

\newcommand{\Xt}{{\tilde X}}

\newcommand{\Yt}{{\tilde Y}}
\newcommand{\Kt}{{\tilde K}}

\newcommand{\et}{{\tilde \varepsilon}}

\newcommand{\om}{{\boldsymbol{\omega}}}

\newcommand{\brh}{{\boldsymbol{\rho}}}
\newcommand{\blam}{{\boldsymbol{\lambda}}}

\newcommand{\La}{\Lambda}

\DeclareMathOperator{\Vect}{\mathrm{Vect}}

\begin{document}

\begin{abstract}
The well-known AKSZ construction (for Alexandrov--Kontsevich--Schwarz--Zaboronsky) gives an odd symplectic structure on a space of maps together with a functional $S$  that  is automatically a solution for the  classical master equation $(S,S)=0$. The input data required for the AKSZ construction consist of a volume element on the source space and a symplectic structure of  suitable parity on the target space, both invariant under given  homological vector fields on the source and target. In this note, we show that  the AKSZ setup and their main construction can be naturally recovered from the single requirement that the `difference' vector field arising on the mapping space  be gradient  (or Hamiltonian). This can be seen as a converse statement for that of  AKSZ. We include a discussion of properties of vector fields on  mapping spaces.
\end{abstract}

\maketitle

\section{Introduction}

\subsection{The AKSZ construction}
What is now known as the `AKSZ construction' was introduced by Alexandrov, Kontsevich, Schwarz and Zaboronsky in~\cite{schwarz:aksz}, who applied it to some models of topological field theory.  The AKSZ construction was further elaborated, for example, in~\cite{cattaneo:sigmaaksz} and ~\cite{roytenberg:akszbv}. It is a construction of a particular solution of the classical master equation on a space of fields together with the equation itself (i.e., an odd symplectic structure)  from certain data on the source and target supermanifolds.

More precisely, the AKSZ construction starts from two supermanifolds, $M$ and $N$ (the source and  target, respectively), together with the following \textbf{input} data:
\begin{itemize}
  \item a homological vector field $Q_1$ on $M$\,,
  \item a homological vector field $Q_2$ on $N$\,,
  \item a volume element $\brh=\rho(x)Dx$ on $M$\,,
  \item a symplectic $2$-form $\o$  of parity $q+1$ on $N$\,,
\end{itemize}
where   $\dim M=p|q\,$, and such  that $\brh$ is invariant under $Q_1$ and $\o$ is invariant under $Q_2$.   It follows that the vector field $Q_2$ is locally Hamiltonian with respect to the symplectic structure $\o$, so that locally $i_{Q_2}\o=-dH$  for some $H$, where $\tilde H=q$. (By the tilde we denote the parities of the objects in question.)

Then as an \textbf{output}  the following objects  on the space of maps $\Mapp(M,N)$ are obtained: the $2$-form
\begin{equation}\label{eq.symplform}
    \O[\f,\d\f]=\int\limits_M \!\!Dx\,\rho(x)\;\o\bigl(\f(x),\d\f(x)\bigr)\,
\end{equation}
and the functional
\begin{equation}\label{eq.action}
    S[\f]=\int\limits_M \!\!Dx\,\rho(x)\; \Bigl(Q_1^a(x)\,\der{\f^i}{x^a}\la_i(\f(x))- H(\f(x))\Bigr)\,,
\end{equation}
where $\o=d\la $ (locally), so that the form $\O$ defines an odd symplectic structure     on $\Mapp(M,N)$
and the functional $S$   satisfies the classical master equation
\begin{equation}
    (S,S)=0\,
\end{equation}
with respect to the corresponding odd Poisson bracket. 
(Since $\la$ is defined only locally, $S$ is an example of a `multi-valued functional', see below.)

Here   the mapping space $\Mapp(M,N)$ is considered as an infinite-dimensional supermanifold. Functions $\f^i(x)$, some of  which may be odd, defining a map $\f\co M\to N$ in local coordinate systems on $M$ and $N$, are regarded as `coordinates' on  $\Mapp(M,N)$. A {form} on $\Mapp(M,N)$ is by definition a function on the antitangent bundle $\Pi T\Mapp(M,N)$. The functions $\f^i(x)$ and their variations $\delta\f^i(x)$, to which we ascribe parities opposite to those of $\f^i(x)$, are together  `coordinates' on the infinite-dimensional supermanifold  $\Pi T\Mapp(M,N)$. This usage agrees with the  standard language in field theory and integrable systems. We use $Dx$ (with the capital $D$) as the notation  for the  Berezin volume element.\footnote{We avoid the notation   `$dx$' for a volume element   because of the   contradictions with the transformation law  under a change of coordinates. The capital letter in $Dx$ should not be confused with the notation for a path integral measure element, e.g.,  in $\mathcal D\f$.}

Note that the functional~\eqref{eq.action}, known as the AKSZ action, is multi-valued, but its variation is well-defined. The study of such multi-valued functionals     was initiated by Novikov in the early 1980s, see,  e.g.~\cite{novikov:multi82}. Formulas such as~\eqref{eq.symplform} for a symplectic  structure  are known. To them  correspond what are known as ``ultra-local'' field-theoretic Poisson brackets, see~\cite{novikov:indynsys4}.

The AKSZ construction has found numerous applications. In the original paper~\cite{schwarz:aksz}, the authors applied it to the Chern-Simons model of topological quantum field theory and to some other models. It was applied to deformation quantization by Cattaneo and Felder~\cite{cattaneo:sigmaaksz}.

\subsection{The main claim. Structure of the paper}
In this note, we show that the AKSZ construction follows naturally from a very  simple setup.

Recall that, given two vector fields $X$ and $Y$ on  (super)manifolds $M$ and $N$, there is a construction of  an induced vector field  on the mapping space $\Mapp(M,N)$. We  denote it by $d(X,Y)$ and call it the \emph{difference construction} for $X$ and $Y$. At each point $\f\in \Mapp(M,N)$, the value of $d(X,Y)$ is defined as the difference $Y\circ \f-d\f\circ X$, which  measures the failure of $X$ and $Y$ to be $\f$-connected.\footnote{Considering  $Y\circ \f-d\f\circ X$ as a tangent vector to the space of maps at $\f$ is  undoubtedly classical, but treating it as a vector field with  variable $\f$, which is a certain shift of view,   probably belongs to~\cite{schwarz:aksz}, though might have appeared earlier.}  Suppose we take  homological vector fields $Q_1\in \Vect(M)$ and $Q_2\in \Vect(N)$. 

\emph{We  claim that the single requirement that the vector field $d(Q_1,Q_2)$ on $\Mapp(M,N)$   be ``gradient'' or ``Hamiltonian'', i.e.,   come from the variation of some action, makes it possible to recover  the whole AKSZ setting, including  formulas~\eqref{eq.symplform} and \eqref{eq.action} for the symplectic form $\O$ and the master action $S$.}

In more detail, this goes as follows.

The master equation $(S,S)=0$ for an action $S$ is equivalent to   the corresponding Hamiltonian vector field $X_S$ being   homological. 
In~\cite{schwarz:aksz}, assuming the whole AKSZ setup described above, it was shown that  the vector field  $X_S$ corresponding to the AKSZ action~\eqref{eq.action} is precisely  the difference construction $d(Q_1,Q_2)$ for the homological vector fields $Q_1$ and $Q_2$. One can see that the difference construction  for  homological fields is automatically homological. Therefore the AKSZ action satisfies the master equation. This was the argument in~\cite{schwarz:aksz}.

In the present paper, we show that    one may start from just two homological vector fields $Q_1$ and $Q_2$ (without initially assuming any other ingredients of the AKSZ scheme) and require simply that the  vector field $d(Q_1,Q_2)$ on   $\Mapp(M,N)$, which is automatically homological, can be written in the gradient form
\begin{equation*}
    d(Q_1,Q_2)= \int_{M} \!\!Dx \;\Psi^{ij}(x,\f(x)) \var{S}{\f^j(x)}\var{ }{\f^i(x)}\,,
\end{equation*}
for some functional $S[\f]$, where no a priori  properties such as symmetry or Jacobi identity are assumed for the object $\Psi^{ij}(x,y)$. Then it turns out that $\Psi^{ij}(x,y)$ automatically comes from an odd symplectic structure on $\Mapp(M,N)$ and we recover uniquely all the AKSZ formulas (in a slightly generalized form). Therefore our construction can be viewed as a `converse' to the AKSZ statement.

The structure of the paper is as follows.

In Section~\ref{sec.vectfields}    we review some general facts about vector fields on spaces of maps. Most of them are known, but we felt that it would be useful to have them assembled together.

In Section~\ref{sec.main} we explain our main construction.

Throughout this note, we follow notations and conventions concerning supermanifolds, homological vector fields and even or odd Poisson brackets that can be found, e.g.,  in~\cite{tv:graded} and \cite{tv:qman-mack}.

\section{General facts about vector fields on mapping spaces}\label{sec.vectfields}

\subsection{Tangent vectors and vector fields on a mapping space}
Consider the space of maps $\Mapp(M,N)$\,. Its ``points'' are smooth maps $\f\co M\to N$.  A \emph{tangent vector} $K$ at $\f$ is an infinitesimal shift $\f\mapsto \f_{\e}=\f+ {\e}K$, i.e.,
\begin{equation*}
    \f^i(x) \mapsto \f^i(x) + \e K^i(x) \quad (\e^2=0).
\end{equation*}
Hence $K$ is a `vector field along the map $\f$\,', i.e., a map $M\to TN$ that covers the  map $\f\co M\to N$ with respect to the projection $TN\to N$\,. As for ordinary manifolds, a tangent vector $K$ to a mapping space $\Mapp(M,N)$   can  be identified with its infinitesimal action on function(al)s $\delta_K$, the   variation  along $K$;   by definition,
\begin{equation*}
    S[\f+\e K]=S[\f]+\e\, \delta_KS \quad (\e^2=0).
\end{equation*}
Here $\delta_K$ is a linear operator mapping functionals to numbers.
The  familiar expansion
 \begin{equation*}
    K=\delta_K=(-1)^{q\tilde K}\int_M\!\! Dx\, K^i(x)\,\var{}{\f^i(x)}\,,
 \end{equation*}
may be regarded as the definition of variational derivatives $\var{}{\f^i(x)}$. Here $\dim M=p|q$. Note, incidentally, that the parity of $\lvar{}{\f^i(x)}$ is $\itt+q$ (not $\itt$). The role of  the sign factor $(-1)^{q\tilde K}$ is in keeping the  linearity of   $\delta_K S$ in $K$ with respect to  the multiplication of  $K$ by   odd scalars.

Hence, a \emph{vector field} $K$ on $\Mapp(M,N)$
gives   infinitesimal shifts for arbitrary maps $\f\in \Mapp(M,N)$,
\begin{equation*}
    \f^i(x) \mapsto \f^i(x) + \e K^i[x |\, \f] \quad (\e^2=0)\,.
\end{equation*}
It is a functional on $\Mapp(M,N)$ taking values in tangent vectors, so that the value $K[\f]$ at $\f$ is a tangent vector at $\f$.
A vector field $K$
can be identified with the corresponding variation $\delta_K$, which is now an operator taking functionals to functionals:
\begin{equation*}
    K[\f]=(-1)^{q\tilde K}\int_M\!\! Dx\, K^i[x|\, \f]\,\var{}{\f^i(x)}\,
 \end{equation*}
and
\begin{equation*}
    \delta_KS\,[\f]=(-1)^{q\tilde K}\int_M\!\! Dx\, K^i[x|\, \f]\,\var{S[\f]}{\f^i(x)}\,,
 \end{equation*}
for a functional $S$.
This is similar to writing vector fields on ordinary manifolds or supermanifolds as differential operators on functions.

The Lie bracket of vector fields on $\Mapp(M,N)$ is defined in the usual way. One  either starts from the group commutator of the infinitesimal diffeomorphisms of   $\Mapp(M,N)$, so that
\begin{equation*}
    1+\e\h\;[K_1,K_2]=(1+\h K_2)^{-1}(1+\e K_1)^{-1}(1+\h K_2)(1+\e K_1)\,,
\end{equation*}
or takes the (graded) commutator of the variations:
\begin{equation*}
    \d_{[K_1,K_2]}=[\d_{K_1},\d_{K_2}]=\d_{K_1}\d_{K_2}-(-1)^{\Kt_1\Kt_2}\d_{K_2}\d_{K_1}\,.
\end{equation*}
In coordinates,
\begin{align*}
    [K_1,K_2] & =   \int_M \int_M Dx Dy \left((-1)^{q\Kt_2}K_1^j[y|\f]\,\var{K_2^i[x|\f]}{\f^j(y)}-
    (-1)^{\Kt_1\Kt_2+q\Kt_1}K_2^j[y|\f]\,\var{K_1^i[x|\f]}{\f^j(y)}\right)\var{}{\f^i(x)}\\
              & =   \int_M Dx \left((-1)^{q\Kt_1}\d_{K_1}K_2^i[x|\f]-(-1)^{\Kt_1\Kt_2+q\Kt_2}\d_{K_2}K_1^i[x|\f]\right)\var{}{\f^i(x)}\,.
\end{align*}

One can imagine various classes of vector fields on a mapping space corresponding to various types of the dependance of  the components $K^i[x| \f]$   on $\f$.   A particular case: $K^i[x| \f]=K^i(x,\p\f(x), \p^2\f(x), \dots, \p^{s}\f(x))$, i.e., the components $K^i[x| \f]$ are differential functions of $\f$ at the same point $x$. Such vector fields $K$ on $\Mapp(M,N)$ are known as \emph{local vector fields}. (The `evolutionary vector fields' of the jet space formalism, cf. Olver~\cite{olver}, were made to mimic exactly this class of vector fields on mapping spaces.) Local vector fields are closed under   commutator.

\begin{rem} Everything above is completely standard at least in the case when $M$ and $N$ are ordinary manifolds. Our goal was mainly to recall the terminology and introduce the notation.
\end{rem}

\begin{rem} The concept of the mapping space $\Mapp(M,N)$ when $M$ and $N$ are  supermanifolds and, in particular, its treatment as an infinite-dimensional supermanifold requires some comments. There are two aspects, the infinite dimensionality and being `super', which are independent of each other. First, even for ordinary manifolds $M$ and $N$, it has to be explained in which sense the set $\Map(M,N)$ of all smooth maps from  $M$ to   $N$  can be itself understood as an `infinite-dimensional manifold'. We refer, for example, to   book~\cite{kriegl:michor-conven} for one particular   approach. In this paper, we follow the `naive' or `formal' viewpoint   used by physicists and do not go into foundations. Secondly, concerning the supermanifold aspect, the subtlety is unrelated with the   infinite-dimensionality. Note that in some cases, for  supermanifolds $M$ and $N$, the mapping space $\Mapp(M,N)$ can be finite-dimensional. (For example, such is the mapping space $\Mapp(\R{0|1},N)$ for any (super)manifold $M$, which coincides with the supermanifold $\Pi TN$; in fact, such are the mapping spaces $\Mapp(\R{0|k},N)$ for arbitrary $k$. In general, $\Mapp(M,N)$ may contain finite-dimensional subspaces that are supermanifolds.) The key fact is that the mapping space $\Mapp(M,N)$ should be regarded as more than just a set consisting of   maps  and   endowed with whatever structure.\footnote{We use boldface to distinguish $\Mapp(M,N)$ from such a set, which we denote $\Map(M,N)$. The set $\Map(M,N)$ with suitable topology is the underlying topological space for $\Mapp(M,N)$. If $M$ and $N$ are ordinary manifolds, there is no need for such a distinction between $\Map $ and $\Mapp $.} Informally,  `odd parameters' should be allowed for these  maps;  and, in the general case, these odd parameters are functional.  To avoid the discussion of an underlying topology and   structure sheaf  for the space $\Mapp(M,N)$, the `convenient formula'
\begin{equation*}
    \Map(P,\Mapp(M,N))=\Map(P\times M,N)
\end{equation*}
may be postulated as a working definition. Here $P$, $M$ and $N$ are supermanifolds, and for fixed $M$ and $N$, and varying $P$, the r.h.s. serves as the definition of the l.h.s. as a functor of $P$. The set  $\Map(P,\Mapp(M,N))=\Map(P\times M,N)$ is, by definition, the set of all $P$-points of the supermanifold $\Mapp(M,N)$. It should be noted that whenever we refer to ``points'' of $\Mapp(M,N)$ or use set-theoretic notation, we always understand points in this generalized sense.
\end{rem}

\subsection{Induced vector fields and the difference construction}

Diffeomorphisms of the source and target induce diffeomorphisms of the mapping space. For $F\in \Diff (M)$ and $G\in \Diff(N)$, we have the transformations $F^*$  and $G_*$ of $\Mapp(M,N)$,
\begin{equation*}
    F^*[\f]=\f\circ F\,,\quad G_*[\f]=G\circ \f\,,
\end{equation*}
which are the usual pull-back and push-forward of a map. Clearly,
\begin{equation*}
    (F_1\circ F_2)^*=F_2^*\circ F_1^*\,, \quad (G_1\circ G_2)_*={G_1}_*\circ {G_2}_*\,.
\end{equation*}

The infinitesimal version of that holds for vector fields.
For  a map $\f\in \Mapp(M,N)$, vector fields on the source and target define its infinitesimal variations. Let $X\in \Vect(M)$ and let $Y\in \Vect(N)$.  We can define vector fields $X^*$ and $Y_*$ on $\Mapp(M,N)$ by the formulas $X^*[\f]:=d\f\circ X$  and $Y_*[\f]:=Y\circ \f$, having in mind the pull-back and the push-forward of $\f$ by the corresponding infinitesimal diffeomorphims:
\begin{align*}
    \f+\e\, X^*[\f]&\, =\, (1_M+\e X)^* [\f]\,=\, \f\circ (1_M+\e X)\,,\\
    \f+\e\, Y_*[\f]&\,=\,(1_N+\e Y)_* [\f] \,=\, (1_N+\e Y)\circ \f\,.
\end{align*}

From the definitions follow the coordinate descriptions:
\begin{equation*}
    X^*=(-1)^{q\Xt}\int_M\!\! Dx \,  X^a(x)\,\p_a\f^i(x)\;\var{}{\f^i(x)}\,
 \end{equation*}
and
 \begin{equation*}
    Y_*=(-1)^{q\Yt}\int_M\!\! Dx \,  Y^i(\f(x))\;\var{}{\f^i(x)}\,.
 \end{equation*}
In particular, both $X^*$ and $Y_*$  are local vector fields.

\begin{prop}
\label{prop.comm}
For arbitrary $X_1, X_2\in \Vect(M)$,
 \begin{equation*}
    [X_1, X_2]^*=-[X_1^*, X_2^*]\,.
 \end{equation*}
 For arbitrary $Y_1, Y_2\in \Vect(N)$,
 \begin{equation*}
    [Y_1, Y_2]_*=[Y_{1*}, {Y_2}_*]\,.
 \end{equation*}
 For arbitrary $X\in \Vect(M)$ and $Y\in \Vect(N)$,
 \begin{equation*}
    [X^*, Y_*]=0\,.
 \end{equation*}
 \end{prop}
The statements are  obvious from the interpretation in terms of  the infinitesimal diffeomorphisms. In particular, the commutativity of $X^*$ and $Y_*$ follows from the commutativity of the left and right shifts.

\begin{cor} If $Q_1\in\Vect{(M)}$ and $Q_2\in\Vect{(N)}$ are homological vector fields, then
the induced vector fields on $\Mapp(M,N)$ are also homological:
\begin{equation*}
    (Q_1^*)^2=0\,,\quad ({Q_2}_*)^2=0\,.
\end{equation*}
\end{cor}

For vector fields $X_1\in \Vect (M)$ and $X_2\in \Vect (N)$ of the same parity, define their \emph{difference construction}, notation: $d(X_1,X_2)$, as the vector field on $\Mapp(M,N)$
\begin{equation*}
    d(X_1,X_2):= X_{2*}-X_1^*\,.
\end{equation*}
Or, equivalently,
\begin{equation*}
    d(X_1,X_2)[\f]:=X_2\circ \f-d\f\circ X_1\,.
\end{equation*}
The zeros of the vector field $d(X_1,X_2)$ are precisely such $\f$ that $X_1$ and $X_2$ are $\f$-related.

\begin{cor} \label{cor.homol}
If vector fields $Q_1\in \Vect (M)$ and $Q_2\in \Vect (N)$ are homological, then the vector field $d(Q_1,Q_2)$ on $\Mapp(M,N)$ is also homological.
\end{cor}
Indeed, $Q_{2*}$ and $Q_1^*$ commute and are homological. Therefore their difference is homological.

\begin{rem} The notion of the difference construction (without such terminology) as a vector field on the space of maps and crucial Corollary~\ref{cor.homol} is due to~\cite{schwarz:aksz}.
\end{rem}

The difference construction has nice properties. For example, suppose $M_1$, $M_2$ and $M_3$
are endowed with vector fields $X_1$, $X_2$ and $X_3$, resp. For any diagram
\begin{equation*}
    \begin{CD}M_1 @>{\f}>> M_2 @>{\psi}>> M_3\,, \end{CD}\,,
\end{equation*}
one may ask about the relation between the vector fields $d(X_1,X_3)$, $d(X_1,X_2)$ and $d(X_2,X_3)$.
\begin{prop} The following identity holds:
\begin{equation}\label{eq.compos}
    d(X_1,X_3)[\psi\circ \f]=d(X_2,X_3)[\psi]\circ \f+d\psi\circ d(X_1,X_2)[\f]\,.
\end{equation}
\end{prop}
The proof is straightforward.

\begin{rem} For homological vector fields, the difference construction should be compared with the familiar definition of the differential on $\Hom(K,L)$ for (co)chain complexes $K$ and $L$. Equation~\eqref{eq.compos} should be compared with the Leibniz rule for this differential with respect to the composition of homomorphisms, to which it reduces in the case of complexes. Notions introduced in this section should be regarded as the non-linear analogs of the corresponding linear notions for complexes. (One can associate  a $Q$-manifold  to a cochain complex, so that the differential becomes a homological vector field, linear in coordinates.)
\end{rem}

Equation~\eqref{eq.compos} and similar identities can be used for introducing the important notion of a \emph{$Q$-category}, generalizing the notion of a \emph{$Q$-group}~\cite{tv:esi-lect}. A \emph{$Q$-category} is a  smooth category such that the morphism `sets' and possibly the `set' of objects are $Q$-manifolds, maybe infinite-dimensional, and all the structure maps are $Q$-morphisms. As an example one can consider some  category of $Q$-manifolds (morphisms --- arbitrary smooth maps). It can be regarded as a \emph{$Q$-category } with respect to the homological vector field $d(Q_{\a},Q_{\b})$ for each $(M_{\a}, Q_{\a})$ and $(M_{\b}, Q_{\b})$. In particular, each supergroup of diffeomorphisms $\Diff M$ for a $Q$-manifold $M$ is a \emph{$Q$-group}.

\smallskip
As we already mentioned in the Introduction, there is an observation (AKSZ): the Hamiltonian vector field $X_S$ corresponding to the AKSZ action is the difference construction for $Q_1$ and $Q_2$. (This explains $(S,S)=0$.)

\textbf{Question:} is it possible, in general, to express  the difference construction for arbitrary vector fields in a ``Hamiltonian'' or ``gradient'' form?

We shall deal with that in the next section.

\section{Main construction and a proof}\label{sec.main}

\subsection{Main statement} Consider supermanifolds $M$ and $N$ and vector fields   $X_1\in \Vect(M)$ and $X_2\in \Vect(N)$ of the same parity $\et$. Here we do not need them to be homological or anything. Therefore our analysis makes sense in the case of ordinary manifolds as well. In the previous section we introduced the difference construction $X_{12}=d(X_1,X_2)$   as a vector field on the mapping space $\Mapp(M,N)$. In coordinates,
\begin{equation}\label{eq.difconst}
    X_{12}  =(-1)^{q\et}\int_M Dx \left(X_2^i(\f(x))-X_1^a(x)\der{\f^i}{x^a}\right)\var{}{\f^i(x)}\,.
\end{equation}
We would like to ask a general question: is it possible to express this vector field on $\Mapp(M,N)$ in a ``gradient form''? That means, is it possible to find a functional $S=S_{12}$ on $\Mapp(M,N)$ and coefficients $\Psi^{ij}(x,y)$ so that\footnote{More precisely, it is a ``local gradient form'' because the arguments in the integrand are taken at the same point $x\in M$.  Note, incidentally, that we are bit sloppy with the common sign   in ~\eqref{eq.gradform} because   it is not important here.}
\begin{equation}\label{eq.gradform}
    X_{12}=\pm\int_M Dx\, \Psi^{ij}(x,\f(x))\,\var{S_{12}}{\f^j(x)}\, \var{}{\f^i(x)}\ ?
\end{equation}
Clearly, the answer depends on the particular $X_1$ and $X_2$, but we would like to find a ``universal'' construction that would work for ``general'' $X_1$ and $X_2$. The functional $S_{12}$ should depend on $X_1$ and $X_2$, while the object $\Psi^{ij}(x,y)$, not. Of course this  means the existence of a certain structure on the manifolds $M$ and $N$. Our task is to identify  this structure.

As we shall see, for this construction to hold,  the vector fields $X_1$ and $X_2$, and the object $\Psi^{ij}(x,y)$ should obey certain constraints.
We shall find them now.

First of all, let us realize what sort of geometric object $\Psi^{ij}(x,y)$ is. It lives on $M\times N$ and carries tensor indices from $N$. Recall that the variational derivative  $\lvar{}{\f^i(x)}$ transforms as a covector  with respect to   $N$ and as a density of weight $1$ (the component of a volume form) with respect to   $M$. We shall apply the terminology such as ``source density'' and ``target covector'', and similar. Then $\Psi^{ij}(x,y)$ is a target tensor of rank $2$ and a source density of weight $-1$ (so that to compensate the total weight of the two variational derivatives).

Note that no symmetry condition in the tensor indices is  assumed a priori for    $\Psi^{ij}(x,y)$.

In a more invariant parlance, objects on   $M\times N$ which we describe by their ``source'' and ``target`` properties are sections of the tensor products of the pull-backs of natural bundles over $M$ and $N$ to the manifold $M\times N$.
Since we can differentiate  objects on $M\times N$   independently in the $M$- and $N$-directions,   it makes   sense to speak, for example,  of a target
$2$-form   on   $M\times N$ as a  target symplectic form, regardless of its source properties.

\begin{thm}  In the setup described above, the difference construction $d(X_1,X_2)$ has a ``universal gradient form''~\eqref{eq.gradform} if and only if the following holds.

The matrix $\|\Psi^{ij}(x,y)\|$ is the inverse of   $\|\o_{ij}(x,y)\|$ and the object $\om$  on $M\times N$ so defined is a target symplectic
form and source density     of weight $1$.

The condition
\begin{equation}
    (L_{X_1}+L_{X_2})\om=0\label{eq.integrabil}
\end{equation}
is satisfied and the action $S_{12}=S_{12}[\f, X_1, X_2]$ is given by the formula
\begin{equation}\label{eq.ouraction}
    S_{12}[\f, X_1, X_2]= \int_M \langle d\f\circ X_1, d_2^{-1}\om \rangle\,+\,d_2^{-1}\left(L_{X_1}d_2^{-1}\om+i_{X_2}\om\right)\,
\end{equation}
(here $d_1$ and $d_2$ denote the exterior differentials on $M$ and $N$, resp.).
\end{thm}

The statement of the theorem deserves a few comments. (A sketch of a proof will follow in the next subsection.) The statement that $\om$ is a target symplectic form means that it is a target $2$-form (with the invertible matrix) and  $d_2\om=0$. It is possible  to apply to $\om$ the Lie derivatives with respect to vector fields on $M$ and $N$. The Lie derivative $L_{X_1}$ acts on $\om$ as on a volume form, while $L_{X_2}$ acts on $\om$ as on a differential $2$-form. Note also that $\om$ defines a symplectic structure on $\Mapp(M,N)$ by the formula
\begin{equation}\label{eq.symplform2}
    \O[\f,\d\f]=\int\limits_M \!\!\om\bigl(\f(x),\d\f(x)\bigr)\,
\end{equation}
similar to~\eqref{eq.symplform}.

As  is clear from the presence of the inverses of the exterior differentials, equation~\eqref{eq.ouraction} defines a functional which is multi-valued. It will be seen from the proof that its variation is well-defined and each $d^{-1}$ in the formula makes sense at least locally. We can check here that $d_2^{-1}$ in the second term makes sense. Indeed, we need to check the $d_2$-closedness; we have $d_2\left(L_{X_1}d_2^{-1}\om+i_{X_2}\om\right)=L_{X_1}d_2d_2^{-1}\om+d_2i_{X_2}\om=L_{X_1}\om+L_{X_2}\om=0$, by ~\eqref{eq.integrabil}.

Let us give a coordinate expression for the action $S_{12}$, which may be useful together with the coordinate-free formula~\eqref{eq.ouraction}. For simplicity we shall write the formulas in the purely even case. Everything extends without effort to the general super case. Suppose  $\om=d_2\blam$ locally. Then $\blam$ is a target $1$-form and source density. In local coordinates, $\blam=Dx\otimes dy^i\la_i(x,y)$ and $\o_{ij}=\p_i\la_j-\p_j\la_i$. Then  we have
\begin{equation}\label{eq.ouractioncoor}
    S_{12}[\f, X_1,X_2]= \int_M Dx\, \left(X_1^a(x)\,\der{\f^i}{x^a}\,\la_i\bigl(x,\f(x)\bigr)\,+U\bigl(x,\f(x)\bigr)\right)\,,
\end{equation}
where the ``potential'' $U(x,y)$ is a target scalar and source density. It is defined from the equation $\p_j U(x,y)= \p_a(\la_j  X^a_1)+X^i_2\o_{ij}$ (the integrability condition for it is exactly~\eqref{eq.integrabil})\,.
\begin{coro}If $\om=\brh\, \o$, where $\brh=Dx \rho(x)$ is a volume element on $M$ and $\o$ is a symplectic form on $N$, then the condition $(L_{X_1}+L_{X_2})\om=0$ becomes $L_{X_1}\brh=0$ and $L_{X_2}\o=0$ and we recover the AKSZ-type setup (for $X_1$, $X_2$ of arbitrary, but equal parity). The action $S_{12}$ takes the form
\begin{equation*}
    S_{12}[\f, X_1,X_2]=\int_M \!\!\brh\; \Bigl(X_1^a\,\der{\f^i}{x^a}\la_i(\f(x))- H(\f(x))\Bigr)\,.
\end{equation*}
Here $\la_i=\la_i(y)\,$,  $d\la=\o\,$, and  $i_{X_2}\o=-dH$.
\end{coro}

\subsection{Sketch of a proof}
For the simplicity of notation consider the purely even case (extending to the general super case is straightforward). We are given that
\begin{equation*}
    X_{12} = \int_M \Psi^{ij}(x,\f(x))\,\var{S_{12}}{\f^j(x)}\, \var{}{\f^i(x)}\,.
\end{equation*}
Since $ X_{12}^i=X^i(y)-X^a(x)y^i_a$, the action $S=S_{12}$  can contain  only   first derivatives and should have the form
\begin{equation*}
    S = \int_M Dx\, \left(\La_i^a(x,\f(x))\,\der{\f^i}{x^a}+U\bigl(x,\f(x)\bigr)\right)\,,
\end{equation*}
so the Lagrangian is $L=\La_i^a(x,y)\,y^i_a+U(x,y)$\,. Calculating the variational derivative $\lvar{S_{12}}{\f^j(x)}$ we obtain
\begin{multline*}
     \der{L}{y^j}-\oder{}{x^a}\left(\der{L}{y^j_a}\right)=
    \der{\La^a_k}{y^j}\,y^k_a + \der{U}{y^j}-\oder{}{x^a}\left(\La_j^a\right)=\\
    \der{\La^a_k}{y^j}\,y^k_a + \der{U}{y^j}-\der{\La_j^a}{x^a}-\der{\La_j^a}{y^k}y^k_a=\\
    \bigl(\p_j\La_k^a-\p_k\La_j^a\bigr)\,y^k_a + \der{U}{y^j}-\der{\La_j^a}{x^a}\,.
\end{multline*}
We should have
\begin{multline*}
    \Psi^{ij}(x,y)\Bigl(\bigl(\p_j\La_k^a-\p_k\La_j^a\bigr)\,y^k_a+\p_jU-\p_a\La_j^a\Bigr)=\\
    -X^a(x)y^i_a+X^i(y)= -X^a(x)\delta^i_ky^k_a+X^i(y)\,.
\end{multline*}
We arrive at the system
\begin{align*}
    \Psi^{ij}\,\bigl(\p_j\La_k^a-\p_k\La_j^a\bigr)&= -X^a\delta^i_k\,,\\
    \Psi^{ij}\,\bigl(\p_jU-\p_a\La_j^a\bigr)&=  X^i\,.
\end{align*}
Note that $\Psi^{ij}(x,y)$ should be universal and not depend on $X_1$, $X_2$, while $\La^a_k$ and $U$ should depend on  $X_1$, $X_2$ by universal formulas. Therefore we conclude, from the first equation,  that $\Psi^{ij}=\Psi^{ij}(x,y)$ is invertible. Introduce the inverse matrix $\o_{ij}$. We obtain
\begin{equation*}
    \p_j\La_k^a-\p_k\La_j^a=-\o_{jk}X^a\,.
\end{equation*}
Therefore $\o_{ij}$ is skew-symmetric. Likewise, we see that $\o_{jk}=\p_j\la_k-\p_k\la_j$ for some $\la_i$. We conclude that $\La_i^a=-\la_iX^a+\p_if^a$, where $f^a=f^a(x,y)$. Hence the Lagrangian is
\begin{equation*}
    L=\bigl(-\la_i(x,y)X^a(x)+\p_if^a(x,y)\bigr)\,y^i_a + U(x,y)\,.
\end{equation*}
Note that $\p_if^a(x,y)\,y^i_a=D_af^a-\p_af^a$, where $D_a$ denotes total derivative with respect to $x^a$. Hence we can pass to an equivalent Lagrangian and re-define $U$ by absorbing $-\p_af^a$\,:
\begin{equation*}
    L= -\la_i(x,y)X^a(x)\,y^i_a + U(x,y)\,.
\end{equation*}
We now look at the second equation from the system above. It gives, after contracting it with $\o_{ki}$, the equation
\begin{equation*}
    \p_kU+\p_a\bigl(\la_kX^a\bigr)= \o_{ki}X^i
\end{equation*}
or
\begin{equation*}
    d_2 U+\p_a\bigl(\blam X^a\bigr)=-i_{X_2}\om  \ \Rightarrow \  \p_a\bigl(\o X^a\bigr)=-L_{X_2}\om \,.
\end{equation*}
Note, finally, that $\p_a\bigl(\om X^a\bigr)=L_{X_1}\om$. Hence we arrive at the relation
\begin{equation*}
    L_{X_1}\om+ L_{X_2}\om=0
\end{equation*}
(this is a Lie derivative of an object on $M\times N$ with respect to the vector field $X_1+X_2$). This is a necessary (and locally sufficient) condition for recovering $U$ in the Lagrangian. For the action, we have arrived at the expression~\eqref{eq.ouractioncoor}, up to an inessential common sign. This concludes the proof. \qed

\section{Examples and discussion}

Applications of the AKSZ construction are numerous. The following examples are recalled   for illustration only.

\begin{ex}[see~\cite{schwarz:aksz}]\label{ex.first}
Consider a supermanifold $M$ of dimension $n|m$. We can take $\Pi TM$ with the de Rham differential $d$ as the homological vector field. This will be the source. The vector field $d$ preserves the canonical volume form $D(x,dx)$ on  $\Pi TM$.   For the target,  consider a symplectic supermanifold $N$ with the symplectic form $\o$ of parity $n+m$. Suppose $H$ is a function on $N$ of parity $n+m+1$ satisfying $(H,H)=0$ with respect to the Poisson bracket generated by the symplectic structure. Let $\o$ on $N$ be locally $d\la$. Note that the parity of $\la$ is $n+m+1$, which is the same as for $H$. On the mapping space $\Mapp(\Pi TM, N)$ we obtain the AKSZ action
\begin{equation}\label{eq.action1}
    S[\f,H]=\int_{\Pi TM} D(x,dx) \left(dx^a\der{\f^i}{x^a}\,\la_i(\f(x,dx))-H(\f(x,dx))\right)\,.
\end{equation}
It satisfies $(S,S)=0$ with respect to the odd symplectic structure on $\Mapp(\Pi TM, N)$ given by the odd $2$-form
\begin{equation*}
    \O[\f,\d\f]=\int_{\Pi TM}\! D(x,dx)\; \o(\f(x,dx),\d\f(x,dx))\,.
\end{equation*}
The Hamiltonian vector field of $S$ with respect to this structure is the difference construction for $d$ and the Hamiltonian vector field $Q=X_H$ on $N$.
\end{ex}

\begin{rem}
In the above example, maps $\Pi TM\to N$ can be interpreted as ``$N$-valued forms'' on $M$. Also, any such map $\f$ naturally lifts to a unique map $\Pi TM\to \Pi TN$ commuting with $d$, which we denote by the same letter $\f$. This allows us to consider pull-backs of forms on $N$ to forms on $M$ with respect to $\f$. Therefore, the functional given by~\eqref{eq.action1} may be re-written simply as
\begin{equation}\label{eq.action2}
    S[\f,H]=\int_M \f^*(\la - H)\,,
\end{equation}
where in the r.h.s. we have the integral of a form\,\footnote{In general, it is a pseudodifferential form, i.e., a not necessarily fiberwise polynomial function on $\Pi TM$. Even for ordinary manifolds $M$ and $N$, the map $\f$ does not have to preserve the degrees of forms, so the integrand in~\eqref{eq.action2} is an inhomogeneous differential form.} over the (super)manifold $M$.
\end{rem}

The following example is a particular case of Example~\ref{ex.first}.

\begin{ex}[Cattaneo and Felder~\cite{cattaneo:felder2000,cattaneo:sigmaaksz}] (It is convenient to change notation slightly.) Consider the space of maps $\Pi TD\to \Pi T^*M$, where $D=D^2$ is a $2$-disk with boundary, and $M$ is a Poisson manifold with an even bracket.  It is specified by a function $P$ on $\Pi T^*M$ (the Poisson bivector), which is fiberwise quadratic and satisfies $(P,P)=0$ with respect to the canonical odd Poisson bracket on $\Pi T^*M$  given by the  canonical odd symplectic form $\o=-dx^adx^*_a=d(dx^a\,x^*_a)$. The de Rham differential as a vector field on $\Pi TD$  preserves the canonical volume form $\brh=D(u,du)$, where $u^i$ are coordinates on the disk.
The homological vector field  $X_P$ on $\Pi T^*M$ which is the odd Hamiltonian field corresponding to the even function $P=\frac{1}{2}\,P^{ab}x^*_b x^*_a$ is nothing but the Lichnerowicz differential for the Poisson cohomology of $(M,P)$.
The AKSZ action written as in~\eqref{eq.action2} is
\begin{equation}\label{eq.action3}
    S[\f,P]=\int_{D^2} \f^*\!\!\left(dx^ax^*_a-\frac{1}{2}\,P^{ab}(x)x^*_b x^*_a\right)\,.
\end{equation}
It turns out that the mapping space $\Mapp(\Pi TD, \Pi T^*M)$ plays the role of the extended phase space of the Batalin--Vilkovisky method.  Namely, one starts from  the space of the vector bundle maps $\Homm(\Pi TD, \Pi T^*M)\subset \Mapp(\Pi TD, \Pi T^*M)$. It turns out that  here
\begin{equation*}
    \Mapp(\Pi TD, \Pi T^*M)\cong \Pi T^*\left(\Homm(\Pi TD, \Pi T^*M)\right)
\end{equation*}
and the AKSZ action~\eqref{eq.action3} on the full space  $\Mapp(\Pi TD, \Pi T^*M)$ plays the role of the Batalin--Vilkovisky extended action with respect to to the same action restricted to the subspace of vector bundle morphisms $\Homm(\Pi TD, \Pi T^*M)$.
Using this method, Cattaneo and Felder showed how to obtain Kontsevich's formulas for deformation quantization of a Poisson manifold $(M,P)$.
\end{ex}

\begin{rem} As mentioned, the above examples are known and are given here just as illustrations. It would be interesting to obtain an example where, as in the previous section, there is no given factorization $\om=\brh \otimes\o$ and objects naturally live on $M\times N$. This may happen if one replaces maps $M\to N$ by sections of   a fiber bundle and $M\times N$ by the total space. (Cf. Kotov and Strobl~\cite{kotov:strobl2007})
\end{rem}

\begin{rem} In our setup for a `gradient form' of the difference vector field given by equation~\eqref{eq.gradform}, we assume that the coefficients  $\Psi^{ij}$ have the form $\Psi^{ij}(ij)(x,y)$, where $y=\f(x)$, i.e.,  depend only on fields, but not  on their derivatives. Allowing a dependence on derivatives in this setting will give a generalization of the AKSZ construction.
\end{rem}

\par\bigskip\noindent
\textbf{Acknowledgement.}
{\small
I have discussed various aspects related with the subject of this note  on several occasions  at the Geometry Seminar in Manchester and at conferences in Lausanne, Lyon, Vienna and Bia{\l}owie\.{z}a. I am grateful to V.~Buchstaber, H.~Khudaverdian, Y.~Kosmann-Schwarzbach, A.~Kotov, S.~Lyakhovich, K.~Mackenzie, M.~Mulase, D.~Roytenberg,  V.~Rubtsov and Th.~Strobl for comments and discussions, and to the organizers of the meetings for the stimulating atmosphere. Special thanks are due to J.~Stasheff, who read the first version of this text and provided numerous helpful comments.

}


\def\cprime{$'$}

\end{document}